\pdfoutput=1

\documentclass[twocolumn,showpacs,superscriptaddress,groupedaddress,floatfix]{revtex4-1}

\usepackage{color}
\usepackage[]{graphicx} 
\usepackage{bm}       
\usepackage{amssymb}
\usepackage{amsmath}  
\usepackage{float}
\usepackage[english]{babel}

\def\per2m{\mathrm{m^{-2}}}
\def\persec{\mathrm{sec^{-1}}}

\def\perm{\mathrm{m^{-1}}}

\def\2D{\mathrm{2D}}
\def\3D{\mathrm{3D}}

\def\Wtrap{\Omega_\mathrm{t}}
\def\ntrap{\nu_\mathrm{t}}

\def\Vout{V_\mathrm{out}}
\def\Vm{V_\mathrm{moving}}
\def\Vf{V_\mathrm{fixed}}
\def\Vdc{V_\mathrm{dc}}

\def\N2{N$_2$}

\begin{document}

\title{A dual-trap system for the study of charged rotating graphene nanoplatelets in high vacuum}
\author{Joyce E. Coppock}
\email{jec@umd.edu}
%\thanks{These authors contributed equally to this work.}
\affiliation{Department of Physics, University of Maryland, College Park, MD}
\author{Pavel Nagornykh}
\thanks{Current affiliation: Center for Nonlinear Dynamics, Department of Physics, University of Texas, Austin, TX}
\affiliation{Department of Physics, University of Maryland, College Park, MD}
\author{Jacob P. J. Murphy}
\affiliation{Department of Physics, University of Maryland, College Park, MD}
\author{I. S. McAdams}
\thanks{Current affiliation: Case Western Reserve University, Department of Electrical Engineering and Computer Science}
\affiliation{Department of Physics, University of Maryland, College Park, MD}
\author{Saimouli Katragadda}
\affiliation{Department of Aerospace, Aeronautical, and Astronautical Engineering, University of Maryland, College Park, MD}
%\thanks{Current affiliation:}
\author{B. E. Kane}
\affiliation{Joint Quantum Institute, University of Maryland, College Park, MD}
\affiliation{Laboratory for Physical Sciences, University of Maryland, College Park, MD}
\date{\today}
\begin{abstract}
We discuss the design and implementation of a system for generating charged multilayer graphene nanoplatelets and introducing a nanoplatelet into a quadrupole ion trap in high vacuum.  Levitation decouples the platelet from its environment and enables sensitive mechanical and magnetic measurements.  The platelets are generated via liquid exfoliation of graphite pellets and charged via electrospray ionization.  A single platelet is trapped at a pressure of several hundred millitorr and transferred to a trap in a second chamber, which is pumped to UHV pressures for further study. 
%[45 words left till abstract limit]
\end{abstract}

\maketitle

\section{Introduction}
The properties of graphene and other two-dimensional materials are strongly affected by the substrate to which they are attached\citep{Chen2008}\citep{Neugebauer2009}, making these materials ideal candidates for levitated study.  Previous work has shown that a charged, micron-scale graphene platelet can be confined in an electric quadrupole trap in high vacuum and caused to rotate at high frequencies using circularly polarized light\citep{Kane2010}.
More recently, we have performed magnetic measurements of unparalleled sensitivity on trapped platelets that have been gyroscopically stabilized using a radio frequency electric field\citep{Nagornykh2016}.  This technique shows promise for other new measurements; for example, a levitated, gyroscopically stabilized nanoplatelet could be incorporated into cavity optomechanics experiments.  Additionally, flexural modes of the rotating, centrifugally tensioned membrane could be studied.

To achieve the high rotational frequencies and torque sensitivities needed for this research, it is necessary to perform the experiments at pressures of $10^{-8}$ Torr and lower.  Since the process of injecting the platelet into the trap generates water vapor and may introduce other contaminants, it is difficult to achieve ultra high vacuum (UHV) pressures in the chamber in which the particle is initially trapped.  To resolve this problem, we perform the initial particle collection using a movable trap, and the particle is subsequently transferred to a trap in an adjacent chamber, which is kept as clean as possible. This paper describes improvements in the procedure for generating and trapping graphene platelets, as well as the mechanics of transferring the particle between traps. 

\begin{figure*}
\begin{center}
\includegraphics[scale=1,draft=false]{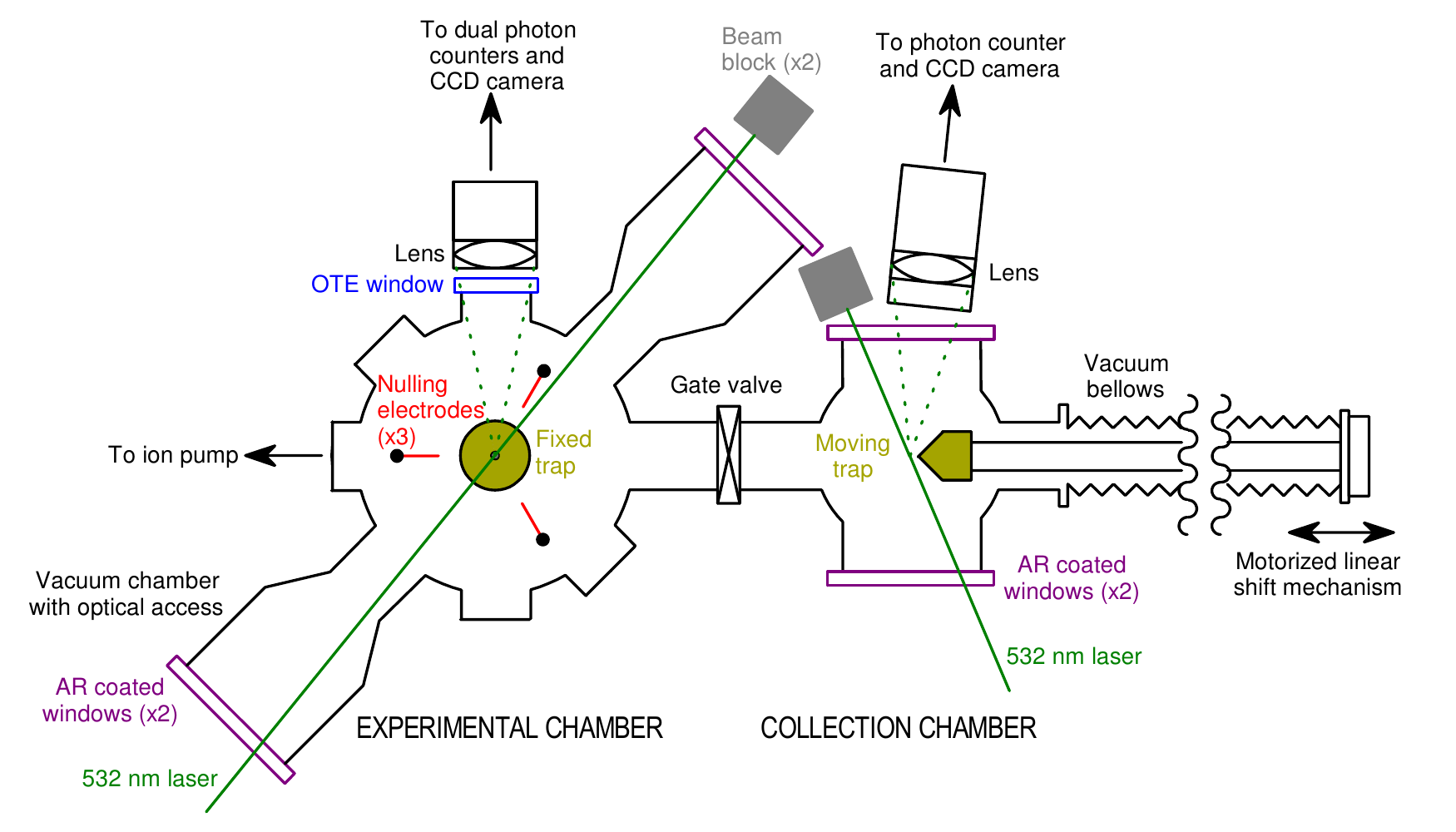} 
\end{center}
\caption{Top view of experimental apparatus.  Particle collection and experimentation are performed in separate chambers to avoid contamination.  Particle is initially collected in a trap mounted on a motorized linear shift mechanism.  With gate valve open, moving trap can be brought near fixed trap and particle can be transferred between traps.  The moving trap is shown fully retracted; in this position the distance between traps is 24.5 cm.  Both chambers may be evacuated by a turbo pump (not shown) or an ion pump.}
\label{fig:transferapparatus}
\end{figure*}

\section{Trap description} 
\label{sec:trap}

\begin{figure}
\begin{center}
\includegraphics[scale=0.7,draft=false]{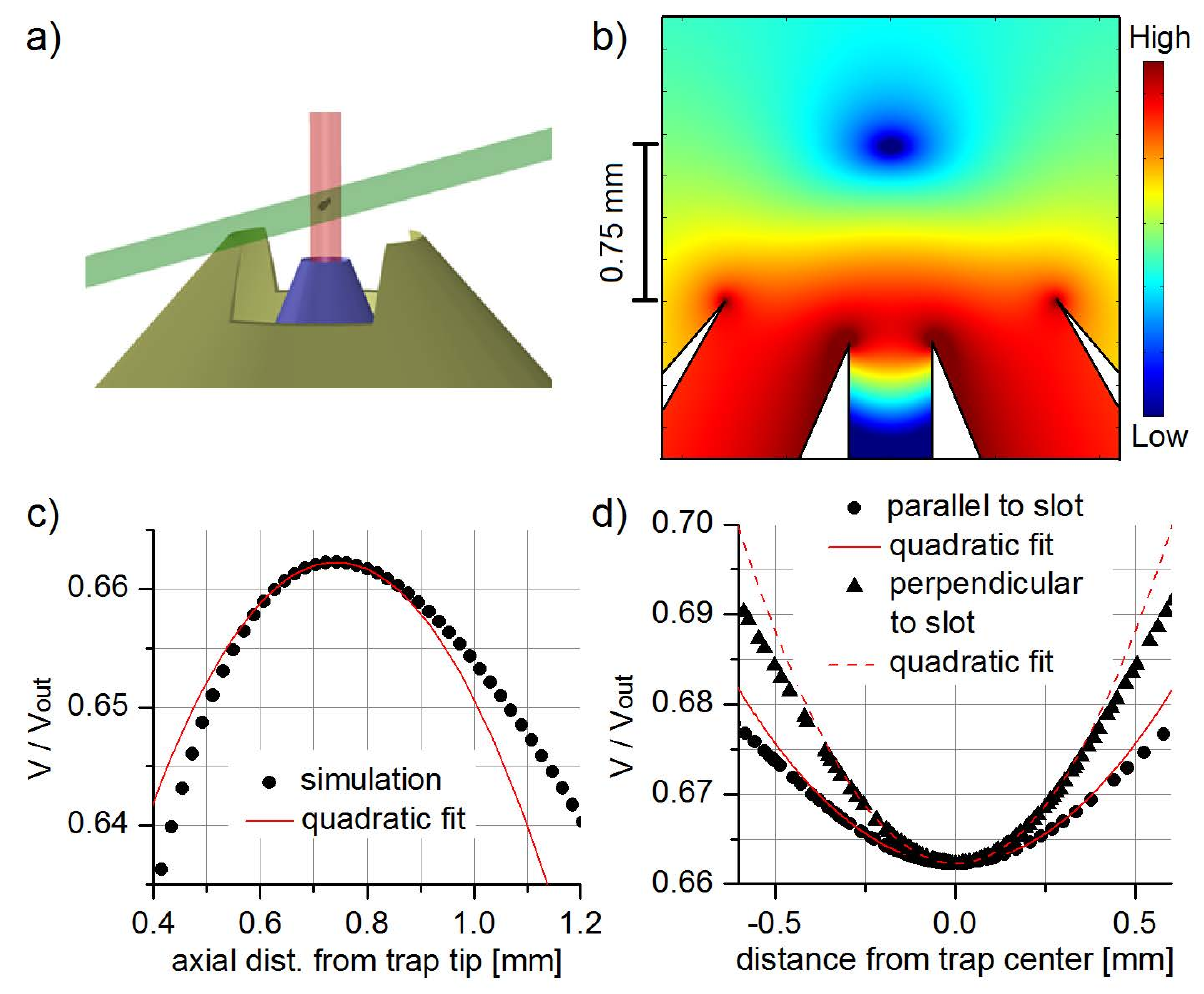} 
\end{center}
\caption{(a) Diagram of tip of trap.  Outer electrode tip diameter 1.6 mm.  Slit width 1.2 mm.  Slit depth 0.6 mm.  Inner electrode tip is recessed 0.2 mm from outer tip.  A hole is drilled through the axis of the inner electrode to admit a circularly polarized 671 nm laser beam (red), which may be used to impart angular momentum to the trapped graphene platelet.  The platelet is imaged via scattering of a 532 nm laser beam (green), directed horizontally.  Platelet shown is not drawn to scale. (b) Cross sectional plot of $|\mathbf{E}|$.  The square of this quantity is proportional to the effective trapping potential given by Eq.\:\ref{eq:pseudopotential}.  A minimum is observed about 0.75 mm from the tip of the trap electrodes.  Section plane is vertical and perpendicular to slot across tip of outer electrode. (c) Plot of electric potential along axis of trap, with voltage $\Vout$ on outer electrode.  Trap center lies at potential maximum, about 0.75 mm from tip of outer electrode. (d) Plot of electric potential in plane normal to axis along lines through trap center, parallel to slot (circles) and perpendicular to slot (triangles). }
\label{fig:backtrap}
\end{figure}

We use traps similar in principle to the Paul trap\citep{Paul1990}, which is commonly used for ion trapping but is suitable for trapping charged micron-sized particles as well.  A stylus geometry\citep{Maiwald2009} allows two traps to be brought very close together, facilitating the transfer of particles between traps.  In this section, we will describe the primary trap used for experiments (the ``fixed trap" in Fig.\:\ref{fig:transferapparatus}); the movable trap used for particle collection is of almost identical design.  

A drawing of the trap geometry is shown in Fig.\:\ref{fig:backtrap}a.  The trap consists of two coaxial conical electrodes made of stainless steel.  The outer electrode has a radius of 0.8 mm at the tip.  The inner electrode has a hole of radius 0.2 mm drilled through its axis to admit a laser beam which is used to impart angular momentum to the particle.  The tip of the inner electrode is recessed by 0.2 mm from the outer electrode tip.  A slot of width 1.2 mm is cut across the tip of the outer electrode in order to break the axial symmetry of the trap.   The trap is enclosed in a vacuum chamber whose walls are grounded.  

%A grounding plane made of stainless steel surrounds the electrodes.  

If a dc voltage of amplitude $\Vout$ is applied to the outer electrode, while the inner electrode is held at ground, the electric field \textbf{E} reaches a zero value at a point $\mathbf{r}=0$ approximately 0.75 mm from the tip of the outer electrode, along the axis of the trap.  An electrostatic potential cannot have a local minimum in three dimensions, so a charged particle cannot be trapped in this configuration.

If, however, the potential applied to the outer electrode varies sinusoidally in time, with $V\left(t\right)=\Vout\cos\left(\Wtrap t\right)$, the particle experiences a pseudopotential given by\citep{Kane2010}:
\begin{align}
\label{eq:pseudopotential}
\Psi\left(\mathbf{r}\right)=\frac{1}{4}\frac{q}{m}\frac{1}{\Wtrap^2}\mathbf{E}^2\left(\mathbf{r}\right),
\end{align}
where $q$ is the charge on the particle, $m$ is its mass, $\Vout$ is the amplitude of the voltage applied to the outer electrode, and $\Wtrap$ is the frequency of $\Vout$.  In the above equation, the damping of the particle's motion due to the background gas has been neglected because experiments are generally conducted at low pressures, where the velocity damping rate is several orders of magnitude smaller than $\Wtrap$.  As the damping rate becomes comparable to $\Wtrap$, the pseudopotential becomes smaller and the particle is more weakly confined.  A color plot of $|\mathbf{E}|$ is shown in Fig.\:\ref{fig:backtrap}b.  The effective potential well is centered at the zero point of the electric field.

Near the point $\mathbf{r}=0$ where \textbf{E}=0, the electric potential may be approximated as a quadratic function:
\begin{align} 
\frac{V\left(x,y,z\right)}{\Vout} =\frac{\alpha_x x^2+\alpha_y y^2+\alpha_z z^2}{2 z_0^2}, \label{eq:V} 
\end{align}
The $z$-axis lies along the axis of the trap electrodes, the $x$-axis lies perpendicular to the slot in the outer electrode, and the $y$-axis lies parallel to the slot.  $z_0$ is a parameter that depends on the electrode configuration.
Plots of the potential along the cartesian dimensions were obtained via electrostatic simulations (using COMSOL \textit{v}4.3a) and are shown in Fig.\:\ref{fig:backtrap}(c-d).  Quadratic fits give $\alpha_x / \alpha_z =0.32$, $\alpha_y / \alpha_z =0.60$, $z_0=1.7$ mm.  The particle is less tightly confined in the direction parallel to the slot.

For small amplitudes of motion, the particle undergoes simple harmonic motion in the well, with characteristic frequencies of oscillation given by\citep{Kane2010}\citep{Nagornykh2015}:
\begin{align} 
\omega_{x,y,z} =\left| \alpha_{x,y,z} \right| \frac{1}{\sqrt{2}}\frac{q}{m}\frac{\Vout}{\Wtrap z_{0}^{2}}, \label{eq:xyzeigenf} 
\end{align}
The presence of three distinct eigenfrequencies allows us to resolve the translational motion of the particle in three dimensions using scattered light collected in a single lens\citep{Nagornykh2015}.
For a recently studied particle with a charge-to-mass ratio of 6.1 C/kg, using a trap frequency of $\ntrap = \Wtrap / \left(2\pi\right)=15$ kHz, we observed eigenfrequencies of $\nu_x=300$ Hz, $\nu_y=450$ Hz, and $\nu_z=750$ Hz, where $\nu_{x,y,z}=\omega_{x,y,z}/\left(2 \pi \right)$.  The observed eigenfrequency ratios $\nu_x / \nu_z =0.4$ and $\nu_y / \nu_z =0.6$ are in good agreement with the ratios predicted by simulation.

The trap is enclosed in a vacuum chamber of diameter 15 cm.  Windows with anti-reflective (AR) coatings permit a laser beam to enter and exit the chamber.  Static electrical charge may build up on the insulating windows; where possible, extra flanges are used to set windows away from the particle to minimize the effects of these charges on the particle.  The window through which the scattered light exits could not be recessed due to the focal length of the lens; to minimize charge buildup, this window is coated with indium tin oxide (ITO) so that it acts as an optically transparent electrode (OTE).

The particle is detected by means of a vertically polarized 532 nm laser directed horizontally across the chamber.  Light scattered from the particle is collected and focused by a lens outside the chamber.  A non-polarizing beamsplitter divides the focused light equally between a ccd camera\footnote{ProEM 512 (Model No. 7555-0002) thermoelectrically cooled ccd camera.  Resolution 512x512 pixels.} and dual high-speed photodetectors\footnote{Hamamatsu MPPC Module C13366-1350GD. The $\sim$20 ns pulses output by this unit are shortened to $\sim$1 ns to increase frequency response to $>$500 MHz}.  Both the laser and the lens/photodetector/camera assembly are mounted on 3-axis translation stages for precise alignment with the trapped particle.  The portion of the beam that is not scattered exits through another window and is collected in an anti-reflective beam dump.

At chamber pressures in the tens and hundreds of millitorr, the particle's motion is strongly damped by the surrounding gas, and spontaneous escape from the trap is rarely observed; particles have remained in the trap for months.  Lower pressures are required, however, for high rotation frequencies and sensitive torque measurements.  At $P < 10^{-7}$ Torr, trap lifetimes are much shorter, typically on the order of hours.  Consequently, we use a feedback cooling method\citep{Nagornykh2015}\citep{Nagornykh2015th} to stabilize the translational motion of the particle, extending the trap lifetime to weeks. 

Stray dc or quasi-static electric fields may develop in the vicinity of the trap\citep{Berkeland1998}\citep{Harter2014}.  They may arise from buildup of charged material on the electrodes or walls of the vacuum chamber.  The presence of such fields enhances the sensitivity of feedback cooling to noise, limiting the achievable cooling efficiency\citep{Nagornykh2015}.  Also, if stray fields are very strong, they may affect the transfer process.  In order to null the effects of stray fields, we apply dc voltages of up to 15 V on each of three nulling electrodes surrounding the trap (see Fig. \ref{fig:transferapparatus}).  

\section{Graphene suspensions}
\label{sec:suspensions}

Graphene platelets are generated via liquid exfoliation, a process in which bulk layered materials are sonicated in a solvent to cause the layers to separate.  Liquid exfoliation has been performed on a large variety of layered materials\citep{Nicolosi2013}, any of which could in principle be studied in this system.  While some research has found that the addition of a surfactant produces suspensions of higher concentration and greater stability\citep{Lotya2010}\citep{Narayan2015}, we avoid the use of any additives that may leave residue on the platelets after evaporation.  

\begin{figure}
\begin{center}
\includegraphics[scale=0.1,draft=false]{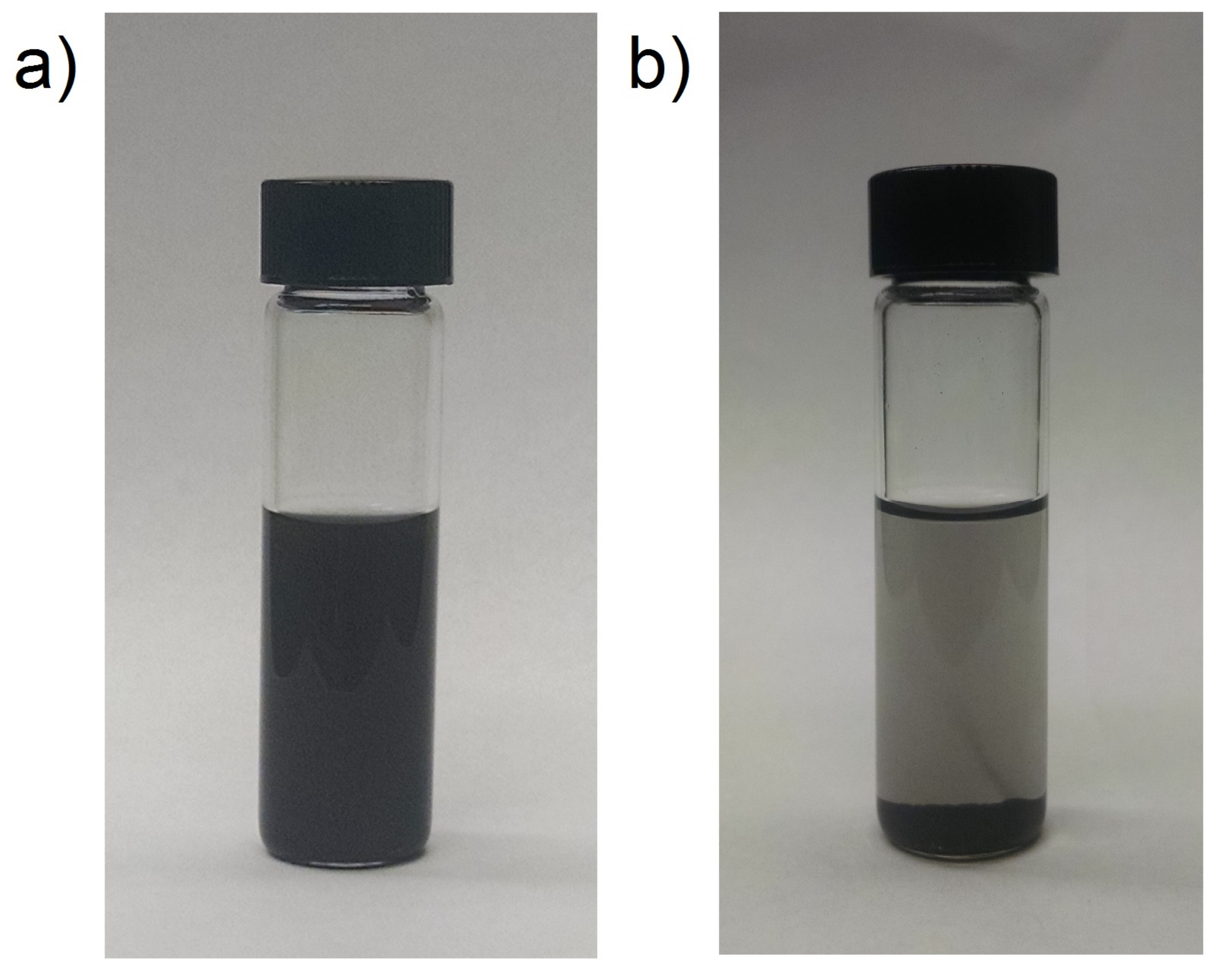} 
\end{center}
\caption{Graphene suspension in isopropyl alcohol water mixture. (a) After sonication.  (b) After centrifugation.}
\label{fig:suspension}
\end{figure}

A brief description of the preparation method follows; details can be found in a previous publication\citep{Kane2010}. 5 mg of graphite pellets are placed in a glass vial with 5 mL of a solution of 3 parts isopropyl alcohol to 1 part deionized water and sonicated for 30 minutes, resulting in a dark liquid containing particles of varying thicknesses, including some unexfoliated graphite chunks (Fig.\:\ref{fig:suspension}a).  This liquid is centrifuged for 30 minutes to bring to the bottom the thickest particles, which sediment the fastest.  The resulting suspension is shown in Fig.\:\ref{fig:suspension}b.  

We characterize the initial suspension density and rate of settling by measuring the attenuation of a 532 nm laser beam directed through the middle of the vial.  The average attenuation coefficient immediately after preparation is 138 $\perm$, with a standard deviation of 33 $\perm$.  Hernandez\citep{Hernandez2008} finds the relationship between concentration and attenuation for exfoliated graphene to be 2500 $\mathrm{L}\,\mathrm{g}^{-1}\,\perm$, which suggests that our suspensions have a concentration of 60 mg\,$\mathrm{L}^{-1}$.  The suspensions settle over time, with the attenuation coefficient dropping approximately exponentially.  The time constant of this settling is on the order of years and increases for suspensions created with fresher alcohol/water solutions.

\section{Particle collection}
\label{sec:collection}
\subsection{Procedure}
\label{subsec:collectionprocedure}

\begin{figure}
\begin{center}
\includegraphics[scale=1,draft=false]{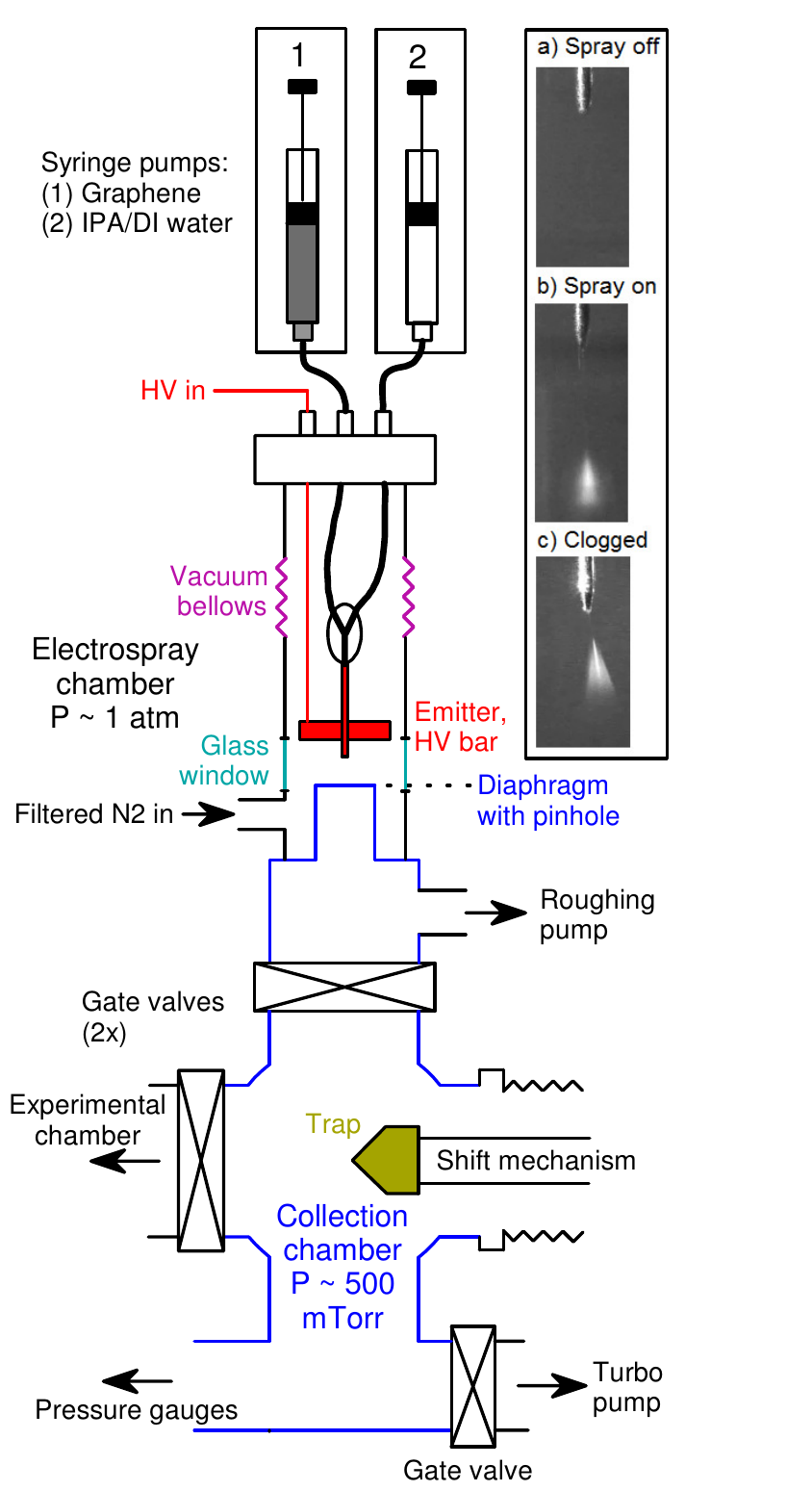} 
\end{center}
\caption{Side view of particle collection chamber.  Drawing is not to scale.  Operation is explained in Sec.\:\ref{sec:collection}.  Syringe pumps are mounted on posts above the electrospray chamber.  The HV bar and emitter are suspended within the electrospray chamber on insulating plastic posts.  The electrospray chamber contains a section of flexible vacuum bellows, which allows the emitter to be positioned precisely over the pinhole.  The top section of the chamber is positioned using a 3-axis translation stage, while the remainder of the apparatus is fixed to the table.  The inset shows close-up photos of the emitter needle illuminated with a laser pointer, with the electrospray (a) off, (b) on, showing an ideal symmetrical cone, and (c) on, showing a skewed cone resulting from a partially clogged emitter.}
\label{fig:collection}
\end{figure}

The suspended graphene particles are given an electrical charge and delivered to the trap via the electrospray ionization technique\citep{Fenn2003}.  The relevant section of the apparatus is shown in Fig.\:\ref{fig:collection}.  Graphene suspension is drawn from the vial using a 22 gauge stainless steel needle into a 500 $\mu$L syringe.  Liquid is injected into the system using a dual syringe pump system\footnote{Chemyx NanoJet}.  One pump holds the syringe of graphene suspension, while the other holds a syringe of pure alcohol/water solution.  The alcohol/water syringe pump is used for cleaning purposes only; during normal particle collection, only the graphene suspension is pumped in.  Tubing made of PEEK (polyether ether ketone, a polymer with good chemical resistance), with an inner diameter of 0.5 mm, is used to carry fluid from each syringe down into a cylindrical chamber (the ``electrospray chamber" in Fig.\:\ref{fig:collection}).  Within the chamber, the tubes empty into a 3-way fitting that directs the liquid flow into a stainless steel needle of inner diameter 100 $\mu$m and length 3.5 cm\footnote{PicoTip TaperTip Emitter, No. MT320-100-3.5-5}.    

%\footnote{Hamilton Gastight No. 1750}

The emitter needle tip sits about 5 mm above a vacuum chamber (the ``collection chamber" in Fig.\:\ref{fig:collection}) whose opening is covered by a stainless steel diaphragm with a 75 $\mu$m diameter pinhole\footnote{National Aperture, Standard Round Aperture, No. 1-75}.  The electrospray chamber column is made with a section of flexible vacuum bellows to allow the emitter needle to be precisely positioned with respect to the pinhole, as well as a glass window to allow viewing of the electrospray cone.  

The electrospray chamber is kept near atmospheric pressure and constantly purged with $\mathrm{N}_2$ gas, while the collection chamber is evacuated by a roughing pump.  The pressure in the collection chamber depends approximately linearly on the diameter of the pinhole, as well as on the pumping rate, and for our system is $\sim$540 mTorr for a clean 75 $\mu$m pinhole.  The emitter needle is held at a voltage of several kV, while the diaphragm is grounded.  The emitter passes through a metal bar with a hole just large enough for clearance; this serves both to keep the emitter vertical and make electrical contact via a wire soldered to the bar.

For liquid flow rates greater than 0.5 $\mu\mathrm{L}\,\mathrm{min}^{-1}$ and emitter voltages of 2100-3000 V, the liquid exits the emitter in a cone of charged microdroplets, some of which contain charged graphene platelets.  This cone can be seen with the naked eye when illuminated with a laser pointer through the glass wall of the electrospray chamber.  Images of the emitter tip when idle and spraying are shown in Fig.\:\ref{fig:collection}(a,b).

The collection chamber holds a movable trap mounted with its axis horizontal. This trap is identical to the one described in Sec.\:\ref{sec:trap}, except that it is axially symmetric, lacking the slit cut across the outer electrode. It also lacks the hole through the axis of the inner electrode.  Its tip is positioned below the pinhole.  A gate valve (the ``electrospray gate valve") lies between the roughing pump inlet and the trap, so that the lower part of the collection chamber can be closed off and evacuated with a turbo pump.  When the gate valve is open, particles can continue their trajectory into the vicinity of the potential well. 

A working set of system parameters for capture of positively charged particles from the suspensions described in Sec.\:\ref{sec:suspensions} are as follows.  A voltage of $+2400$ V is applied to the emitter tip, while liquid is expelled from the emitter at a rate of 1 $\mu\mathrm{L}\,\mathrm{min}^{-1}$.  An oscillating voltage is applied to the outer electrode of the trap, with amplitude $\Vout=300$ V and frequency $\ntrap=35$ kHz.  A dc bias $\Vdc=-3$ V is applied to the inner electrode.  Negatively charged particles can be collected by simply reversing the polarities of the emitter voltage and the dc bias.  

Particles in the collection chamber are detected in a similar manner to those in the experimental chamber: a 532 nm laser beam is directed through the vicinity of the trap and the scattered light is collected by a lens outside the chamber and focused onto a CCD camera and photodetector\footnote{Hamamatsu MPPC Module C12661-150.  Photon detection efficiency is 35\% at 532 nm.} (see Fig.\:\ref{fig:transferapparatus}).  Laser power is set to approximately 1 mW for particle visibility during collection, although smaller powers are used for most experiments.  During electrospraying, particles can typically be seen on the CCD camera passing through the vicinity of the trap at a rate of approximately 1 per second.  Over a few minutes, ten or more particles typically collect in the trap.  Unwanted particles may be selectively expelled from the trap by adjusting the trap frequency $\ntrap$ or the inner electrode dc bias $\Vdc$.  Decreasing $\ntrap$ expels particles with higher charge-to-mass ratios, while increasing $\Vdc$ to a positive value of a few volts (or a negative value if collecting positively charged particles), with $\ntrap$ at a relatively high value, expels particles with lower charge-to-mass ratios.  

For a particle of a given charge-to-mass ratio, there is a minimum trap frequency $\nu_\mathrm{min}$ below which the particle cannot be confined.  As $\ntrap$ approaches $\nu_\mathrm{min}$, the potential well becomes shallow and the image of the particle on the CCD camera changes shape.  For transfer and subsequent study, $\ntrap$ is set to $1.5 \times \nu_\mathrm{min}$.  

\subsection{Collecting silica particles}
\label{subsec:silica}

In addition to graphene platelets, it is possible to capture silica nanoparticles.  We prepared a solution of manufactured silica spheres of diameter 416 nm\footnote{www.microspheres-nanospheres.com, Product No. 147070-10} in the same alcohol/water solution used for graphene suspensions.  The spheres were dispersed in the solution by shaking the vial; they stay suspended for a matter of hours.  The spheres are readily captured using the same voltage and trap frequency settings as are used for graphene.  Their large size makes them useful for calibration of optics and testing of transfer procedures.

\subsection{Cleaning of the system}
\label{subsec:cleaning}

With continued use of the particle collection system, graphene particles build up on the inside of the emitter needle.  When the emitter is clean, the electrospray cone disappears within ten seconds after the syringe pump is turned off; however, as the needle becomes clogged, the stopping time increases.  We have observed that stopping times in excess of 30 seconds are often associated with a decreased rate of trapping of large ($\sim1~\mu$m) particles, suggesting that the built up graphene material may be partially blocking the opening.  Occasionally, long stopping times are also associated with a skewed electrospray cone, as seen in Fig.\:\ref{fig:collection}c.  In extreme cases, the electrospray may refuse to start at all.  

We attempt to forestall graphene buildup by spraying alcohol/water mixture after each session of particle collection for at least 8 minutes, long enough for several flushes of the plumbing downstream of the 3-way fitting.  If a particle is in the collection chamber trap, the electrospray gate valve can be closed during this process.  Despite these precautions, the system still eventually shows signs of clogging, usually after several cumulative hours of electrospray.  The blockage can usually be cleared by removing the syringe pump and emitter assembly from the system and manually expelling 2000-5000 $\mu$L of alcohol/water mixture through the emitter at a fast flow rate.  In some cases, the emitter cannot be cleaned and must be replaced.

Particles also build up around the edge of the pinhole as they pass through, reducing the effective pinhole diameter.  This results in a gradual drop in pressure in the collection chamber with continued electrospraying, and usually also a drop in the rate of collection of large particles.  For a 75 $\mu$m diameter pinhole, the system can be operated for approximately 2-3 cumulative hours of electrospraying before the pressure drops to $\sim$350 mTorr, around which point particles large enough to be convenient for study (on the order of 1 $\mu$m in diameter) are no longer collected and the diaphragm must be replaced.  Smaller pinholes become clogged even faster.

\section{Transfer of particle between traps}
\label{sec:transfer}

\begin{figure*}
\begin{center}
\includegraphics[scale=.09,draft=false]{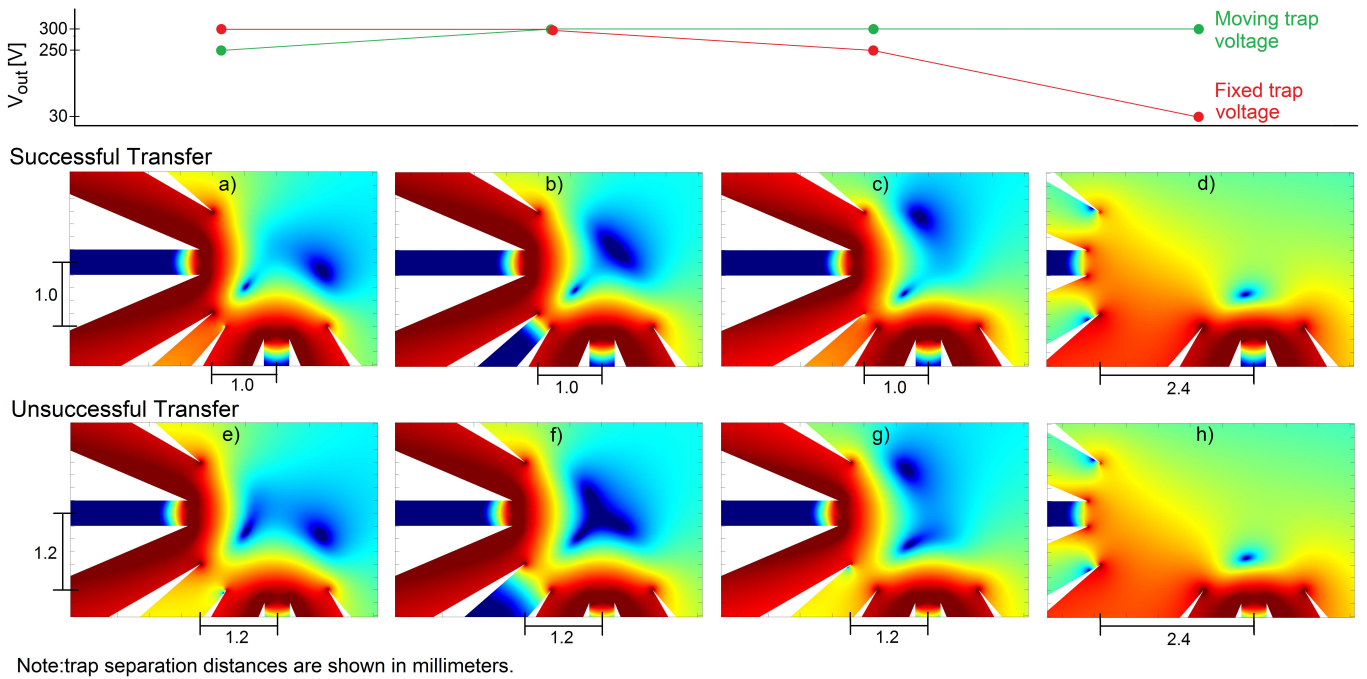} 
\end{center}
\caption{Cross-sectional view of ponderomotive potential landscape at four ``snapshot" moments during transfer.  The section plane is vertical, so that the moving trap appears on the left and the fixed trap appears at the bottom.  The slot in the fixed trap is neglected for simplicity.  The trap voltages for each of the four snapshots are shown at top.  A successful transfer is shown in parts (a-d). (a) With $\Vm=300$ V, $\Vf=250$ V, the traps are brought into proximity such that the tip of each trap is offset by 1 mm from the axis of the other trap.  The particle lies in the lower left hand well.  (b) $\Vf$ is increased to 300 V.  (c) $\Vm$ is decreased to 250 V.  The empty well moves toward the upper left, while the well containing the particle moves toward the fixed trap.  (d) The moving trap is retracted to about 2.4 mm from the axis of the fixed trap as $\Vm$ is decreased to 30 V.  The well containing the particle now lies above the fixed trap.    An unsuccessful transfer is shown in parts (e-h). (e) The tip of each trap is offset by 1.2 mm from the axis of the other trap. (f) As $\Vf$ is increased, the two wells combine.  (g) When the wells separate as $\Vm \to 250$ V, the particle remains in the upper well unless a large dc voltage is applied to the auxiliary electrodes to draw the particle down toward the fixed trap.  (h) As $\Vm \to 30$ V, the upper well disappears and the particle is lost.}
\label{fig:transfersim}
\end{figure*}

A diagram of the two traps is shown in Fig.\:\ref{fig:transferapparatus}.  The trap in which the particle is collected (the "moving trap" in Fig.\:\ref{fig:transferapparatus}) is mounted with its axis of symmetry horizontal on a rod 200 mm in length.  The trap is moved using a linear shift mechanism\footnote{uhvdesign.com, LSM38-300-H}.  The shift mechanism is controlled with the combination of a servo motor\footnote{Teknic, CPM-MCPV-34320-ECN-1-0-1}, and a planetary gearhead\footnote{Parker Hannifin Co., PX34-010-82} with gear ratio 10:1.  A micrometer is fitted to the shift mechanism to read out the position of the trap, which can be controlled to within 0.1 mm.  The trap is shifted with an acceleration of  $\sim$0.8 mm$\,\mathrm{sec}^{-2}$ and a maximum speed of $\sim$0.8 mm$\,\persec$.  Gradual acceleration is used to avoid particle escape.

Initial particle capture is performed with the gate valve between the two chambers closed.  Once a particle is captured, the collection chamber is pumped down to below $1\times10^{-6}$ Torr in order to evacuate most of the water vapor residue.  The gate valve is opened to allow the moving trap to pass through.  At microtorr pressures, the particle's motion is weakly damped, and it is likely that the particle will escape from the trap (as the potential well becomes shallower) during transfer.  For this reason, pressure is increased to 30 millitorr using helium gas from a clean source\footnote{Matheson Ultra High Purity}.  The outer electrode voltages are set at maximum ($\Vm=300$ V) for the moving trap and roughly 80\% of maximum ($\Vf=250$ V) for the fixed trap.  The moving trap moves along its own axis of symmetry, which is positioned about 1 mm above the tip of the fixed trap.  The tip of the moving trap is brought to a distance of about 1 mm from the axis of the fixed trap.  Optionally, the inner electrode dc bias may be set at $\Vdc=-4V$ to pull the particle closer to the trap while it is moved.  Especially for smaller particles, this seems to decrease the likelihood of particle loss.  After movement of the trap, the bias is reset to zero.  The potential landscape at this point in the transfer process is shown in Fig.\:\ref{fig:transfersim}a.  It is usually possible to see the particle on the CCD camera in the subsequent steps of the transfer process.  

Two potential wells are now visible.  The particle sits in the lower left well but has a chance of escaping to the upper right well as the potential barrier between the wells drops during transfer.  With the traps stationary, the voltage on the fixed trap is increased to maximum ($\Vf = 300$ V).  The resulting potential landscape is shown in Fig.\:\ref{fig:transfersim}b. The voltage on the moving trap is then decreased to $\Vm = 250$ V, causing the well containing the particle to move closer to the fixed trap and the empty well to move off to the upper left of the diagram (Fig.\:\ref{fig:transfersim}c).  The moving trap voltage is further decreased to $\Vm = 30$ V in small increments as the trap is retracted to a distance of 2.4 mm (Fig.\:\ref{fig:transfersim}d).  Finally, the moving trap is retracted to its initial position in the collection chamber.  The experimental chamber can now be isolated by closing the gate valve.

Both practice and simulation show that the ideal displacement of the tip of each trap from the axis of the other trap is 1 mm.
A transfer simulation with each trap displaced by 1.2 mm is shown in Fig.\:\ref{fig:transfersim}(e-h).  In this case, the two wells join together (Fig.\:\ref{fig:transfersim}f) as the fixed trap voltage is increased to maximum.  As the moving trap voltage is decreased and the wells separate and move apart, the particle is likely to remain in the upper well and fail to transfer to the fixed trap.  With vertical and horizontal separations ranging from 1.2-1.4 mm, a successful transfer can still sometimes be achieved by applying a large voltage to all three auxiliary nulling electrodes (shown in Fig.\:\ref{fig:transferapparatus}) to pull the particle downwards to keep it in the correct well.  For trap displacements of 1.6 mm or greater, the well containing the particle completely fails to transfer to the fixed trap.  

In the present experimental configuration, the trap voltages are sourced from a dual-channel signal generator and amplified using identical high-voltage amplifiers\footnote{TREK Model 2205}.  During transfer, the ac voltages on the two traps must be in phase with each other.  Experience has shown that a phase difference of more than about 10 degrees will result in an unsuccessful transfer.

\section{Conclusions}
\label{sec:conclusions}
In order to enable experiments on graphene nanoplatelets maximally decoupled from their environment, we have established a method of collecting a nanoplatelet in a quadrupole ion trap and transferring it to a second trap in a chamber that can be pumped down to UHV pressures. Collection of particles large enough for convenient study (with estimated diameters on the order of 1 $\mu$m) is reasonably efficient.  With a clean system, the median time to collection of such a particle is roughly 15 minutes of electrospraying, with standard deviation 10 minutes.  For this size of particle, the transfer success rate at a pressure of 30 mTorr is nearly 100\%, assuming proper alignment of the traps.  Smaller particles are more often lost during transfer; the reason for this is not yet understood, but may be related to the fact that smaller particles typically have larger charge-to mass ratios.  

An area meriting further study is the graphene suspension recipe.  Particles collected have been estimated to have thicknesses $\sim$10 layers or more, while the ideal platelet would be single-layer.  It is possible that a different solvent mixture would achieve better exfoliation of the graphite.
%Also, while an upper limit on the chamber pressure during particle collection are imposed by Paschen's Law,  it is possible that the ideal pressure for particle collection at pressures higher than 550 mTorr has not been attempted.  
Also, further improvements to the transfer procedure may be possible.  While desired experimental pressures are $10^{-8}$ Torr or lower, the current transfer procedure is reliably successful only at pressures of $\sim$10 mTorr and higher.  Fine-tuning of the trap voltages and separation distances may enable transfer at lower pressures, although a limit will be imposed by the fact that the particle tends to escape from the trap at pressures below $\sim$1 $\mu$Torr in the absence of feedback cooling (which cannot be used during transfer).  

Since particles are detected purely optically, measurements of characteristics such as size, shape, and thickness must be performed indirectly and involve considerable uncertainty.  We are developing a method of expelling the flake from the trap and directing its trajectory onto a substrate where it can be characterized more precisely via conventional microscope.  Kuhlicke and colleagues have demonstrated success at a related task, that of depositing silica microspheres from a Paul trap onto an optical fiber\citep{Kuhlicke2015}.

We predict that our graphene suspensions can be replaced with any one of the wide variety of 2D materials that can be generated by liquid exfoliation\citep{Nicolosi2013}.  Our transfer technique could be used to deliver samples to environments inhospitable to the collection process: for example, to traps inside strong magnets for measurements in the quantum Hall regime.

\section{Acknowledgements}

This work was supported by the Laboratory for Physical Sciences.

\bibliography{apparatus2017}

\end{document}